\documentstyle[aps,epsfig,rotating]{revtex}

\def\beq{\begin{equation}}
\def\eeq{\end{equation}}

\def\reff#1{(\ref{#1})}

\def\operator#1{\mbox{\sf #1}}
\def\poperator{\operator{p}}
\def\xoperator{\operator{x}}

\def\hamop{{\operator{H}}}

\def\halb{\frac{1}{2}}

\def\Edach{\hat{E}}
\def\Adach{\hat{A}}
\def\alphadach{\hat{\alpha}}

\def\energy{{\cal{E}}}

\def\pabl#1#2{\frac{\partial #1}{\partial #2}}

\def\ket#1{| #1 \rangle}
\def\braket#1#2{\langle #1 | #2 \rangle}

\def\imagi{\mbox{\rm i}}
\def\diff{\,\mbox{\rm d}}

\begin{document}

\title{Electron correlation vs.\ stabilization: A two-electron model atom\\ in
  an intense laser pulse}
\author{D.~Bauer and F.~Ceccherini}
\address{Theoretical Quantum Electronics (TQE)\cite{www}, Darmstadt University of Technology,\\ Hochschulstr.\ 4A, D-64289 Darmstadt, Germany}
\date{\today}

\maketitle

\begin{abstract}
We study numerically stabilization against ionization of a fully correlated two-electron model
atom in an intense laser pulse.  We concentrate on two frequency regimes: very
high frequency, where the photon energy exceeds both, the ionization potential
of the outer {\em and} the inner electron, and an intermediate frequency
where, from a ``single active electron''-point of view
the outer electron is expected to stabilize but the inner one is not.
Our results reveal that correlation reduces stabilization when compared to results
from single active electron-calculations. However, despite this destabilizing
effect of electron correlation we still observe a decreasing ionization
probability within a certain  intensity domain in the high-frequency case.
We compare our results from the fully correlated
simulations with those from simpler, approximate models. 
This is useful for future
work on ``real'' more-than-one electron atoms, not yet
accessible to numerical {\em ab initio} methods. 
\end{abstract}


\section{Introduction}
The advent of high intensity lasers led to an increasing interest in
non-perturbative studies of atomic systems interacting with intense laser
light (see, e.g., \cite{prot} for a review). One of the most frequently revisited topics during the
last fifteen years was {\em stabilization} of atoms (or ions) against
ionization in intense laser light, i.e., for increasing laser intensity the
ionization rate {\em decreases}. This kind of stabilization was predicted
by Gersten and Mittleman already in 1975 \cite{mittle}. Experimentally, 
stabilization of highly excited atoms has been reported \cite{boer} whereas measuring
stabilization of atoms initially in the ground state is hard to achieve. This is
due to the fact that, in order to see stabilization, the laser photon energy has to exceed the ionization
potential. 
Unfortunately, there are not yet
high intensity lasers available delivering such energetic photons.    
Therefore most of the studies in this field are of analytical or numerical
nature: ``high-frequency theory'' \cite{gavrila}, Floquet calculations \cite{doerr,faisal},
the numerical treatment of 1D model atoms, quantum \cite{su,grobe,grobe_ii,vivi} and classical \cite{rosen},
as well as in two-color laser fields \cite{cheng},
2D atoms in arbitrary polarized laser light \cite{patel}, and full 3D hydrogen \cite{kul}. 
Of particular interest is whether the atom survives in a ``real'' laser pulse up
to intensities where stabilization sets in, or whether it already ionizes almost 100\%
during the rise time of the pulse \cite{lamb}. In other words: is the atom able
to pass through the ``death valley'' of ionization before arriving at the ``magic mountain''
of stabilization? There are also several
papers where the authors came to the conclusion that stabilization does not exist at all (see \cite{fring,geltman}, and references therein).

In this paper we focus on how the electron correlation in a two-electron model
atom affects the probability for stabilization, i.e., the probability that the
model atom remains neutral after the pulse has passed by. 
For two frequency regimes we compare the results from the fully correlated
calculation with approximate models like ``single active electron'' or
time-dependent density functional theory. The purpose of these studies is, on
one hand, to gain a qualitative picture of the stabilization mechanism in a more-than-one
electron atom, and, on the other hand, testing approximate methods before
applying them to 3D many-electron atoms where accurate, full {\em ab initio} studies are
not possible with current days computers. 
To our knowledge only a few other numerical studies of correlated two-electron systems in the
stabilization regime are reported in the literature so far
\cite{grobe_iii,lewen,volkova}.

\section{The model atom} \label{modelatom} 
We study a model helium atom where both electrons are allowed
to move in one dimension only, but with the electron-electron correlation fully
taken into account. This leads to a two-dimensional time-dependent
Schr\"odinger equation (TDSE)
\beq \imagi \pabl{}{t} \ket{\Psi(t)} = \hamop(t) \ket{\Psi(t)} \eeq
with the Hamiltonian
\beq \hamop(t)= \halb( \poperator_1 + A(t))^2 + \halb( \poperator_2 + A(t))^2
- \frac{2}{\sqrt{\xoperator_1^2 + \epsilon}} - \frac{2}{\sqrt{\xoperator_2^2 +
    \epsilon}} +   \frac{1}{\sqrt{(\xoperator_1-\xoperator_2)^2 + \epsilon}}
. \label{full_hamil} \eeq
Here, the laser pulse is coupled to the atom in dipole approximation through the vector potential
$A(t)$. $\xoperator_i$ and $\poperator_i$ ($i=1,2$) are the electrons'
coordinates and canonical momenta, respectively. We use atomic units (a.u.)
throughout this paper. The regularization parameter $\epsilon$ was chosen
$0.49$ which yielded, on our numerical grid,  ionization potentials similar to real helium (0.9 a.u.\
for the first electron, and 2 a.u.\ for the second one).

The electric field $E(t)=-\partial_t A(t)$ was a trapezoidal pulse with a
rising edge over 5 optical cycles, 5 cycles of constant amplitude $\Edach$,
and a down-ramp over, again, 5 cycles. We started all our simulations with the
field-free ground state $\ket{\Psi(0)}$. The wavefunction
$\Psi(x_1,x_2,t)=\braket{x_1 x_2}{\Psi(t)}$ was propagated in time using an
unconditionally stable, explicit ``grid hopping'' algorithm
\cite{raedt}. Non-vanishing probability amplitude $\Psi(x_1,x_2)$ near the grid
boundary was removed through an imaginary potential. 
The numerical grid was always several times (at least 10 times) larger than the excursion length 
\beq \alphadach=\vert \Edach/\omega^2\vert = \vert \Adach / \omega \vert \eeq
of a classical electron oscillating in the laser field of frequency $\omega$
and electric field amplitude $\Edach$ (vector potential amplitude $\Adach$).
During time
propagation we monitored the amount of probability density $\vert\Psi\vert^2$
inside a box $x_1,x_2\in[-5,+5]$. After the pulse is over, the density
inside this box can be interpreted as the ``survival'' probability of the
helium atom to remain neutral \cite{box_comment}.  

To analyze the results obtained with this fully correlated model atom we
compare with several simplified models. Among those, the ``single active
electron'' (SAE) approximation is the simplest one. There, one assumes that an {\em
  inner} and an {\em outer} electron respond independently to the laser
field. The inner electron ``feels'' the bare nucleus ($Z_i=2$, hydrogen-like). The outer one
sees an  effective nuclear charge, to be adjusted in such a way that
the correct ionization potential (0.9~a.u.) is obtained. 
In our numerical model this was the case for $Z_o=1.1$.
Thus, in the SAE approximation, we solved two independent TDSEs with no
dynamic correlation at all, 
\begin{eqnarray}
\imagi\partial_t\Psi_i(x,t) &=& \left(\halb( -\imagi\partial_x + A(t))^2 
- \frac{Z_i}{\sqrt{x^2 + \epsilon}}\right) \Psi_i(x,t), \\
\imagi\partial_t\Psi_o(x,t) &=& \left(\halb( -\imagi\partial_x + A(t))^2 
- \frac{Z_o}{\sqrt{x^2 + \epsilon}}\right) \Psi_o(x,t).
\end{eqnarray}

In order to incorporate correlation in a first step one can introduce a
Hartree-type potential into the Hamiltonian for the inner electron,
\begin{eqnarray}
\imagi\partial_t\Psi_i(x,t) &=& \left(\halb( -\imagi\partial_x + A(t))^2 
- \frac{Z_i}{\sqrt{x^2 + \epsilon}} + \int \frac{\vert
  \Psi_o(x',t)\vert^2}{\sqrt{(x-x')^2 + \epsilon}} \diff x'
\right) \Psi_i(x,t), \label{iso_i} \\
\imagi\partial_t\Psi_o(x,t) &=& \left(\halb( -\imagi\partial_x + A(t))^2 
- \frac{Z_o}{\sqrt{x^2 + \epsilon}}\right) \Psi_o(x,t). \label{iso_ii}
\end{eqnarray}
In this approximation, the inner electron feels the bare nuclear potential
{\em and} the outer electron. Therefore, we call this model ``inner sees
outer'' (ISO) approximation. It was utilized in Ref.~\cite{watson} to study
non-sequential ionization (NSI). In the ground state, the
Hartree-potential leads to a screening of the bare nuclear charge. Thus,
energetically the
two electrons are almost equivalent in the beginning, though we labelled them
``inner'' and ``outer'' in Eqs.\ \reff{iso_i}, \reff{iso_ii}. However, during the
interaction with the laser field one of the electrons might become the outer
one. We will also consider the opposite point of view where the outer electron
sees the inner one (``outer sees inner'', OSI). In this case we have to deal
with the system of TDSEs
\begin{eqnarray}
\imagi\partial_t\Psi_i(x,t) &=& \left(\halb( -\imagi\partial_x + A(t))^2 
- \frac{Z}{\sqrt{x^2 + \epsilon}} 
\right) \Psi_i(x,t), \label{osi_i} \\
\imagi\partial_t\Psi_o(x,t) &=& \left(\halb( -\imagi\partial_x + A(t))^2 
- \frac{Z}{\sqrt{x^2 + \epsilon}}+ \int \frac{\vert
  \Psi_i(x',t)\vert^2}{\sqrt{(x-x')^2 + \epsilon}} \diff x'\right) \Psi_o(x,t). \label{osi_ii}
\end{eqnarray}
with $Z=2$.

Finally, another way to study our model system is to apply time-dependent
density functional theory (TDDFT) (see \cite{gross} for an overview) in local density approximation, leading to
the (nonlinear) TDSE for {\em one} Kohn-Sham orbital $\Phi(x,t)$,
\beq 
\imagi\partial_t\Phi(x,t) = \left(\halb( -\imagi\partial_x + A(t))^2 
- \frac{2}{\sqrt{x^2 + \epsilon}} + \int \frac{\vert
  \Phi(x',t)\vert^2}{\sqrt{(x-x')^2 + \epsilon}} \diff x'
\right) \Phi(x,t). \label{dft} \eeq
The total electron probability density is given by
$n(x,t)=2\vert\Phi(x,t)\vert^2$. Since, strictly speaking,  $\Phi(x,t)$ cannot be
interpreted as a physically meaningful single electron orbital it is
not easy to deduce single electron quantities (such as single ionization for
instance). On the other hand, the somewhat arbitrary distinction between an
inner and an outer electron is avoided in TDDFT.

\section{High-frequency results}
\subsection{Single active electron-approximation} \label{sae_section}
In this Section we want to compare the results from the fully correlated model
atom with those from the corresponding SAE calculations. First, we focus on the
high-frequency regime where the laser frequency exceeds both
ionization potentials, $\omega = \pi > \energy_i >\energy_o$. From an
SAE-point of view we expect both electrons to stabilize, especially the outer
one since the frequency is 3.5 times larger than the ionization potential
$\energy_o=0.9$. In Fig.~\ref{fig_i} the amount of probability density inside the
box  $x_1,x_2\in[-5,+5]$ (PDIB) vs.\ time for $\alphadach=0.5,1.0,1.5$ and $2.0$ is shown
for the inner and the outer electron. First we observe the expected result
that the outer electron is more stabilized than the inner one: the amount of
PDIB after the pulse has gone is greater for the
outer electron. However,
qualitatively the set of curves are quite similar. For the two higher $\alphadach$-values ($1.5$, drawn dashed, and $2.0$, drawn dashed-dotted) one observes that
the curves bend sharply after the up-ramping at $t=5$ cycles, i.e.,
for both electrons ionization is much slower during the constant, intense part of the
pulse. In fact, ionization happens almost exclusively during the rampings.
We also note that the slight decrease of the PDIB during the constant
part of the $\alphadach=1.5$ and $\alphadach=2.0$-pulses is linear in time, in
contrast to tunneling ionization where we have an exponential dependence (in
the case of a constant ionization rate).
Approximately two cycles after the down-ramping at $t=10$ cycles, ionization
starts to increase again. Finally, after the pulse is over (at $t=15$ cycles)
the amount of PDIB remains stationary.
Secondly, we observe that there is obviously no monotonous behavior of
stabilization with increasing intensity (or $\alphadach$). Ionization is higher
for $\alphadach=1.0$ (dotted curves) than for $\alphadach=0.5$ (solid curves). 
For $\alphadach=1.5$ (dashed curves) ionization starts to
decrease, i.e., we are entering the stabilization domain at that
point. Therefore the so-called ``death valley'' seems 
to be located around $\alphadach=1$ for {\em both} electrons in our model atom. 

In Fig.~\ref{fig_ii} the stabilization probability for both
electrons is shown vs.\ the excursion $\alphadach$. 
The quiver amplitude $\alphadach$ is related to the laser
intensity $I=\Edach^2$ through $\alphadach=I^{1/2}\omega^{-2}$.  
The stabilization probability of the inner electron exhibits an oscillatory
behavior. The ``death valley'' is located at $\alphadach\approx 1$,
followed by a maximum at $\alphadach\approx 4$. For higher intensity ionization
increases again up to a stabilization minimum around $\alphadach\approx 7$.
Stabilization of the inner electron recovers till the next maximum at $\alphadach
\approx 9$. The second maximum is below the first. Thus we observe an overall
decrease of stabilization with increasing intensity. This is even more
pronounced in the stabilization probability for the outer electron where the
oscillations are less visible. The ``death valley'' for both electrons is
located at $\alphadach\approx 1$ while the maxima are at different positions.
The oscillatory character of the stabilization probability in 1D systems has
been observed by other authors as well \cite{yao,mill,su_ii}. In contrast to our
results an overall increase of stabilization with increasing intensity was
found in \cite{yao,mill}. This might be due to the fact that we are looking at
the ionization {\em probability} after the pulse is over while in analytical
papers often the ionization {\em rate} is discussed. In the former, ionization
during the up and down-ramps is taken into account while in the latter it is
commonly not.

The probability for our He model atom to survive as neutral He after the
pulse is over is, in SAE approximation, simply the product of the probabilities for each electron to
remain bound. In Fig.~\ref{fig_ii} the corresponding curve is indicated by
i$\cdot$o. The result from the fully correlated system is also shown (drawn
dotted, indicated with `corr'). We infer that, especially for $\alphadach < 5$ the
stabilization probability is strongly overestimated by the SAE treatment. 
We could argue that {\em if} the system stabilizes it will
probably stabilize in such a way that the correlation energy is minimized. In
that case it sounds more reasonable to take the square of the SAE stabilization probability for the
{\em inner} electron. In what follows we will refer to this viewpoint as
``independent electron'' (IE) model since it follows from crossing out the
correlation term in the Hamiltonian \reff{full_hamil}. 
The result is included in Fig.~\ref{fig_ii}, labelled
i$\cdot $i. The IE-curve seems to oscillate around the fully correlated one,
especially for $\alphadach>6$. From
this result we conclude that, compared to the IE-model with two equivalent inner
electrons,  electron correlation washes out oscillations in the stabilization
probability and, therefore, can stabilize
as well as destabilize, depending on the intensity (for a given pulse shape
and frequency).  

To discuss that further we look at the time-averaged Kramers-Henneberger
potential (TAKHP), i.e., we transform to the frame of reference where the
quivering electron is at rest but the nuclear potential oscillates, and average
over one cycle,
\beq V_{KH}^{corr}(x_1,x_2)=\sum_{i=1}^2 \frac{\omega}{2\pi} \int_0^{2\pi/\omega}
\frac{-2}{\sqrt{(x_i+\alpha(t))^2 +0.49}} \diff t + \frac{1}{\sqrt{(x_1-x_2)^2 +
    0.49}} .\label{av_KH} \eeq
For sufficiently high frequencies this is the leading term in a perturbation
series in $\omega^{-1}$ \cite{gavrila}. In the correlation term no $\alpha(t)$ appears
since the interparticle distance is not affected by the KH transformation.
We calculated numerically the TAKHP. The result is shown in Fig.~\ref{fig_iii}
for $\alpha(t)=\alphadach \sin\omega t$ with $\alphadach=5$. For comparison
the TAKHP with the correlation term neglected is also shown (corresponding to
the IE model).
With correlation, there are two minima near $x_1=\alphadach, x_2=-\alphadach$ and
  $x_1=-\alphadach, x_2=\alphadach$ whereas without correlation there are two more,
  energetically equivalent minima at $x_1=x_2=\alphadach$ and $x_1=x_2=-\alphadach$. However,
  if we assume that the fully correlated system manages it {\em somehow} to occupy
  the ground state of $V_{KH}^{corr}$, the correlation energy will be small (for not too small
  $\alphadach$) since the interparticle distance is $2\alphadach$. The higher
  $\alphadach$ the lower the correlation energy. We believe that this is the
  physical reason that, for increasing $\alphadach$, the agreement of the IE
  results with the fully correlated ones becomes quite good (although the
  latter do not exhibit an oscillating stabilization probability). Our
  viewpoint is further supported by examining the probability density of the
  fully correlated system during the pulse. In Fig.~\ref{fig_iv}
  $\vert\Psi(x_1,x_2)\vert^2$ is shown for $\omega=\pi$, $\alphadach=4.0$ at
  $t=7.5$ cycles, i.e., in the middle of the constant part of the trapezoidal pulse.
We clearly observe {\em dichotomy}, i.e., two probability density peaks at the
classical turning points, well known from one-electron systems \cite{gavrila}. Due to
electron correlation we do not observe four peaks. Instead the peaks at
$x_1=x_2=\pm 4$ are suppressed, in accordance with our discussion of the
TAKHPs in Fig.~\ref{fig_iii}. 
Therefore the correlation energy is rather small since the distance between
the two peaks in the $x_1x_2$-plane is $\sqrt{8}\alphadach$.
In the work by Mittleman \cite{mittle_ii} such multi-electron ``dichotomized'' bound
states are calculated.

\subsection{Time-dependent density functional theory} \label{tddft_section}
In Fig.~\ref{fig_v} our results from the TDDFT calculations are
presented. Although the Kohn-Sham orbital $\Phi(x,t)$ is an auxillary entity
that has, in a rigorous sense, no physical meaning, we take it as an
approximation to a single electron orbital. If we do this,
$\vert\Phi(x,t)\vert^2\cdot \vert\Phi(x,t)\vert^2$, integrated over the region
$-5 < x < 5$ after the pulse is over, is our TDDFT stabilization probability.
We see that for $\alphadach<1.5$ the agreement between TDDFT and correct
result is very good. The difference between TDDFT and IE (indicated by
i$\cdot$i again) is a direct measure of correlation effects since both models
differ by the Hartree-term in the Hamiltonian only. Up to $\alphadach\approx
5.5$ electron correlation suppresses stabilization compared to the IE
approximation. In that region TDDFT agrees better with the full result. As
mentioned before, for higher $\alphadach$ the IE curve oscillates around the correct result and
therefore it comes occasionly to a very good agreement with the exact result.
Also the TDDFT result agrees very well with the fully correlated curve for
$\alphadach\geq 7$. In summary we can say that the TDDFT result is in good
agreement with the exact, fully correlated stabilization probability. Both
have their maximum around $\alphadach\approx 4$ and the ``death valley'' is also
at the right position. For higher $\alphadach$ the agreement seems to become
even better. 

\subsection{``Inner sees outer'' and ``outer sees inner''-approximation}
In order to explain non-sequential ionization (NSI; see, e.g., Ref.\ \cite{prot}
for an overview)  it is essential to
incorporate electron correlation (see \cite{becker} for a very recent paper,
and references therein). For that purpose Watson {\em et al.}\ \cite{watson} added
a Hartree-type potential to the TDSE for the inner electron (see Eq.\
\ref{iso_i}, ISO). By doing this the double ionization yield is greatly enhanced,
in accordance with experimental results \cite{walker}. The question we address in this
Section is whether this method is applicable to stabilization
as well. We will also study the opposite procedure, i.e., where the outer electron
feels the inner one (see Eqs.\ \ref{osi_i} and \ref{osi_ii}, OSI). 
From the discussions on the SAE approximation above we
can expect that ISO will probably not agree very well with the exact results
since the assumption that the outer electron sees a static, effective nuclear
charge is not valid. In Fig.~\ref{fig_vi} we see that, for low $\alphadach$
ISO ($\bigtriangleup$) behaves like SAE (the i$\cdot$o-curve) while OSI
($\Diamond$) is similar to IE (the curve indicated by i$\cdot$i). In ISO
approximation for $\alphadach>3$ the electron correlation obviously causes
strong ionization, compared to the SAE result. Especially during the
down-ramping, when probability density of the outer electron moves from the
turning points $\pm\alphadach$ back toward the nucleus, ionization of the
inner electron is enhanced. For $\alphadach>4.5$ the ISO curve even drops
below the exact result (indicated by `corr'). In OSI
approximation the stabilization probability is also underestimated for
$\alphadach>4.5$.
In summary we can say, that for $\alphadach<2$, OSI is in very good agreement
with the correct result while ISO is not, due to the inappropriate assumption
of the outer electron feeling just a static effective nuclear charge. However,
for higher $\alphadach$ ISO and OSI tend to underestimate the stabilization
probability while  TDDFT does not (see Section~\ref{tddft_section} and Fig.\ \ref{fig_v}).

\section{Intermediate frequency results}
In this Section we discuss the stabilization probability in the intermediate
frequency regime $\energy_o < \omega < \energy_i$ where, 
according a single active electron point-of-view, the
outer electron should stabilize while for the inner one ionization is more
likely. In Fig.\ \ref{fig_vii} we compare the result from the fully correlated
calculation with those from the SAE treatment. In SAE approximation, the outer
electron is more stable than the inner one in the region $1.5 < \alphadach <
8.5$. For the inner electron no clear stabilization maximum is visible. For
the outer electron the maximum is at $\alphadach\approx 6$, i.e., it is
shifted toward higher $\alphadach$ compared to the high-frequency case. 
Both, i$\cdot$o and i$\cdot$i underestimates ionization, especially for low
$\alphadach$ in the ``death valley''-region. 
Electron correlation obviously enhances ionization. For lower
frequencies this is the well-known effect of NSI (\cite{becker} and references
therein). Although in the fully correlated result we
observe a stabilization probability maximum around $\alphadach\approx 7.5$ the
absolute value is below $0.04$, and, in our opinion, it makes no sense to talk
about real ``stabilization'' in that case. As in the high-frequency-case we
observe an overall decrease of the stabilization probability for very high $\alphadach$-values.

In Fig.\ \ref{fig_viii} we compare the result from the fully correlated model
atom with the corresponding ones from the ISO, OSI, and TDDFT runs
where electron correlation is included approximately. Let us first focus on
the $\alphadach$-region ``left from the death valley'', i.e., $\alphadach\leq
1$. There we observe that ISO is nearest to the correct result while OSI and
TDDFT underestimates ionization. This is quite understandable within the
present knowledge of how NSI works: the inner electron needs to interact with laser
field {\em and} the outer electron in order to become free. Obviously this is best
accounted for in the ISO approximation. For most $\alphadach$ TDDFT lies
between the ISO and the OSI result. This is also quite clear since in TDDFT both correlated
electrons are treated on an equal footing (one Kohn-Sham-orbital only) whereas
in ISO (OSI) the inner (outer) electron feels the outer (inner) partner
through Coulomb correlation, but not vice verse. However, all these
approximations still overestimate the stabilization probability, at least in
the interesting $\alphadach$-regime where the stabilization probability rises
at all (i.e., for $ 2 < \alphadach < 7.5$). 

To summarize this Section we can say that in order to achieve stabilization of
our two-electron model atom it is necessary to choose a laser frequency that
exceeds {\em all} ionization potentials. For an intermediate frequency the
outer electron cannot stabilize owing to correlation. The SAE picture is not
appropriate and even ISO, OSI, or TDDFT where electron correlation is included
approximately fail.

\section{Discussion and summary}
In this paper we studied how the electron correlation in a two-electron model
atom affects the probability for stabilization. We found clear stabilization only for
frequencies that exceed {\em both} ionization potentials. Although for the
intermediate frequency we did not find a monotonous increase of the ionization
probability with increasing intensity we prefer not calling this effect
stabilization since, on an absolute scale, its probability was very small.
In all cases electron correlation
reduced the stabilization probability compared to the SAE picture. In the
high-frequency case the two electrons behave more like two independent
{\em inner} electrons. Similar results were obtained by Grobe and Eberly
\cite{grobe_iii} for a H$^-$ model-ion. Lewenstein {\em et al.}\ \cite{lewen} 
performed classical calculations for a model-atom similar to ours which also
showed that ``dichotomized'' two-electron states are dynamically accessible.

The agreement of the exact numerical result with TDDFT in the high-frequency case was
quite good while in the intermediate frequency regime stabilization was
overestimated by {\em all} approximate techniques (ISO, OSI, and TDDFT).

It is well-known that a slow time-scale in the stabilization
dynamics is introduced owing to floppings between states in the time-averaged
Kramers-Henneberger potential \cite{gavrila,vivi}. It can be easily imagined that
these slow floppings are affected by electron correlation because, e.g., the
merging of the two dichotomous peaks into a single one is suppressed
then. But even without correlation the results for the stabilization
probability are quite sensitive to rise time and pulse duration of the laser
field since it  strongly depends on which Kramers-Henneberger states are mainly
occupied at the time instant when the laser pulse ends. To avoid these
additional complications in the interpretation of the numerical results we
chose a rather short laser pulse duration so that low-frequency Rabi-floppings
do not play a role. Therefore, in the high-frequency studies we just observed the
two 
dichotomous peaks building up (as
depicted in Fig.~\ref{fig_iv}) but no peak-merging during the constant part of
our trapezoidal laser pulse.

Finally, we would like to comment on the reduced dimensionality of our
two-electron model atom. We also performed calculations with ``real'', i.e.,
three-dimensional (3D) hydrogen-like ions in the stabilization regime. 
It seems to be the case that in 3D stabilization is less pronounced. Moreover,
the oscillatory character is less visible, i.e., we observe a single
stabilization maximum followed by a rather monotonous increase of the
ionization probability. The difference of 1D models and 3D hydrogen was also
studied in Ref.~\cite{mill}. The effect of electron correlation in 3D
stabilization will
be the subject of a future paper \cite{cecch}.

\section*{Acknowledgment}
Fruitful discussions with Prof.\ P.\ Mulser are gratefully acknowledged.
This work was supported in part by the European Commission through the TMR
Network SILASI (Superintense Laser Pulse-Solid Interaction), No.\
ERBFMRX-CT96-0043, and by the Deutsche Forschungsgemeinschaft under Contract
No.\ MU 682/3-1.

\pagebreak

\begin{figure}
\caption{\label{fig_i} The amount of probability density inside the box  
$x_1,x_2\in[-5,+5]$ vs.\ time for $\alphadach=0.5,1.0,1.5$ and $2.0$ (drawn solid,
dotted, dashed, and dashed-dotted, respectively) for the inner and the outer
electron in ``single active electron''-approximation. The laser pulse of frequency $\omega=\pi$ was ramped-up linearly (in field) over 5 cycles, held
constant for another 5 cycles before the linear down-ramp
between $t=10$ and $t=15$~cyc.  As expected, the outer electron is more stabilized. For the two
higher $\alphadach$-values we clearly see that ionization is weak during the
{\em intense, constant} part of the pulse. }
\end{figure}

\begin{figure}
\caption{\label{fig_ii} Stabilization probability for inner (i) and outer (o) electron
  in SAE-approximation for $\omega=\pi$. The fully correlated treatment leads to the
  stabilization probability (i.e., the probability for the He model atom to
  survive as {\em neutral} He) drawn dotted and indicated with `corr'. From a
  SAE viewpoint one would expect the product of the curves i and o, indicated
  with `i$\cdot$o'. Obviously, in SAE approximation the stabilization
  probability is overestimated. The so-called ``death valley'' (dv) for small $\alphadach$ is
  located around $\alphadach\approx 1$ for both, SAE and fully correlated
  results. The i$\cdot$i-curve results from an independent electron model with
  two inner electrons. This curve seems to oscillate around the correct curve
  `corr'.}
\end{figure}

\begin{figure}
\caption{\label{fig_iii} The time-averaged Kramers-Henneberger potential with
  ($V_{KH}^{corr}$) and without ($V_{KH}$) electron correlation included for
  $\alphadach=5$.
With correlation, there are two minima near $x_1=\alphadach, x_2=-\alphadach$ and
  $x_1=-\alphadach, x_2=\alphadach$ whereas without correlation there are two more,
  energetically equivalent minima at $x_1=x_2=\alphadach$ and $x_1=x_2=-\alphadach$. However,
  once the fully correlated system is in the ground state corresponding to
  $V_{KH}^{corr}$, the correlation energy is small (for not too small
  $\alphadach$)
  since the interparticle distance is $2\alphadach$. }
\end{figure}

\begin{figure}
\caption{\label{fig_iv} Probability density $\vert\Psi(x_1,x_2)\vert^2$ for $\omega=\pi$, $\alphadach=4.0$ at
  $t=7.5$ cycles, i.e., in the middle of the constant part of the trapezoidal pulse.
We clearly observe {\em dichotomy}, i.e., two probability density peaks around
$x_1=\mp 4, x_2=\pm 4$. Therefore the correlation energy is rather small since the distance between
the two peaks in the $x_1x_2$-plane is $\sqrt{8}\alphadach$.}
\end{figure}

\begin{figure}
\caption{\label{fig_v} Stabilization probability in the high-frequency case
  ($\omega=\pi$) calculated from time-dependent density functional theory (TDDFT). The
  fully correlated results (labelled `corr') and the independent electron
  curve (indicated by i$\cdot$i) are included for comparison. The agreement of
  TDDFT with the exact result is good: ``death valley'' and stabilization maximum
  are at the same $\alphadach$ position. For higher $\alphadach$ the agreement
  is even better. 
  }
\end{figure}

\begin{figure}
\caption{\label{fig_vi} Comparison of ``inner sees outer'' (ISO, $\bigtriangleup$) and ``outer sees inner'' (OSI, $\Diamond$) results with
  the fully correlated ones (corr). Also shown are the ``single active
  electron'' (SAE) stabilization probability (i$\cdot$o) and the
  ``independent electron'' (IE) prediction (i$\cdot$i), already presented in
  Fig.\ \ref{fig_ii} and discussed in Section \ref{sae_section}. For low $\alphadach$
ISO behaves like SAE while OSI agrees with IE.  For $\alphadach>4.5$ the ISO curve drops
below the exact result. In OSI
approximation the stabilization probability is also underestimated for
$\alphadach>4.5$.}
\end{figure}

\begin{figure}
\caption{\label{fig_vii} Stabilization probability for inner (i) and outer (o) electron
  in SAE-approximation for the intermediate frequency $\omega=\pi/2$. 
The fully correlated treatment leads to the
  stabilization probability drawn dotted and indicated with `corr'. From a
  SAE viewpoint one would expect the product of the curves i and o, indicated
  with `i$\cdot$o'. The i$\cdot$i-curve results from an independent electron model with
  two inner electrons. All these models underestimate ionization, i.e., they
  overestimate stabilization (please note that the stabilization probability is on a
  logarithmic scale now).}
\end{figure}

\begin{figure}
\caption{\label{fig_viii} Comparison of ``inner sees outer'' (ISO,
  $\bigtriangleup$), ``outer sees inner'' (OSI, $\Diamond$), and
  time-dependent density functional theory (TDDFT, $+$) results with
  the fully correlated ones (corr). Stabilization is overestimated in the ISO,
  OSI, or TDDFT approximation.}
\end{figure}


\begin{thebibliography}{99}
\bibitem{box_comment} The atom might be in an excited state after the pulse is
  over. In order to check this and allow for autoionization we continued the
  calculation for another 10 atomic time units (for $\omega=\pi$) or 20~a.u.\
  (for $\omega=\pi/2$).
\end{thebibliography}
\end{document}